\documentclass[referee]{raa}

\usepackage{graphicx,times}
\usepackage{natbib}
\usepackage{amssymb,amsmath}
\bibpunct{(}{)}{;}{a}{}{,}
\usepackage[pagebackref=true]{hyperref}

\begin{document}

\title{Onsite Calibration of the Shear-Shear Correlation}

\volnopage{ {\bf 20XX} Vol.\ {\bf X} No. {\bf XX}, 000--000}
\setcounter{page}{1}

\author{Cong Liu\inst{}, Jun Zhang\inst{*}, Zhenjie Liu\inst{}}
\institute{ State Key Laboratory of Dark Matter Physics, School of Physics and Astronomy, Shanghai Jiao Tong University, Shanghai 200240, People's Republic of China; {\it betajzhang@sjtu.edu.cn}\\
\vs \no
   {\small Received 20XX Month Day; accepted 20XX Month Day}
}

\abstract{Calibration of the cosmic shear bias is crucial for justifying the cosmological results. However, it is still unclear to what extent calibrations based on simulated galaxy images, as what is commonly done in the weak lensing community, can capture the real shear bias, especially given the complicated instrumental effects. On the other hand, selections of the source galaxies (magnitude cut, redshift binning, etc.) made in real measurement, as well as stochastic but possibly correlated shear biases, may introduce errors to the shear-shear correlations that are hard to calibrate apriori. In our previous few papers, we have shown that the field-distortion signal associated with each galaxy image can be saved in the shear catalog to provide onsite calibrations of shear bias for both galaxy-galaxy lensing and shear-shear correlation. In this paper, we apply this method to the HSCpdr3 shear catalog generated by the Fourier\_Quad shear measurement method. Using our onsite calibration method, we find that the shear biases vary with the selections of photo-z bins, SNR cuts, optical bands, as well as alternative forms of the shear estimators. Nevertheless, after calibrations, the shear-shear correlation functions and cosmological parameter constraints show consistent results in all the cases considered. The fiducial results from the r/i/z bands shear catalog with $\mathrm{SNR}>10$ cut yield: $S_8=0.740^{+0.030}_{-0.030}$ and $\Omega_m=0.383^{+0.129}_{-0.132}$.
}

\authorrunning{Liu et al. }            %author_head in even pages
\titlerunning{Onsite Calibration of the Shear-Shear Correlation}  % title_head in odd pages
\maketitle

%________________________________________________ sections below
%
\section{Introduction}

Weak gravitational lensing refers to the small but coherent distortions of the images of distant galaxies by the large scale structure of the universe. Since the first measurement of weak lensing \citep{2000MNRAS.318..625B,2000ApJ...537..555K}, it has been widely used to study the nature of dark matter, dark energy, and gravity \citep{2008ARNPS..58...99H,2017MNRAS.467.4131V,LDP}. One of the most important statistics in weak lensing study is the cosmic shear, namely the two point statistics of background shear. As cosmic shear is directly related to the matter power spectrum of the universe, it provides tight constraints on the cosmological parameters \citep{DESY3cosmos,HSCY3cosmos,KiDS1000cosmos}, especially $S_8=\sigma_8(\Omega_m/0.3)^{0.5}$, which is a combination of the matter density ($\Omega_m$) and its spatial fluctuation amplitude ($\sigma_8$). In recent years, several Stage III surveys take weak lensing as their primary scientific goals, including the Dark Energy Survey (DES, \cite{DESsurvey}), the Kilo-Degree Survey (KiDS, \cite{KiDssurvey}), and the Hyper Suprime-Cam survey (HSC, \cite{HSCsurvey}). These surveys aim to cover thousands of square degrees, providing us huge and high quality imaging data for weak lensing studies.

Weak lensing signals are generally very small (typically at the level of 1\%), and therefore measuring weak lensing shear is challenging. Various systematic errors (e.g. CCD systematics, Point Spread Function (PSF) uncertainties, selection effects) may contaminate the shear measurement, leading to biased cosmological results. Thus, it is crucial to calibrate and control these systematic errors in the weak lensing analysis. So far, major weak lensing surveys all rely on survey-like image simulations to calibrate the shear measurement biases \citep{DES_imsim,KiDS_imsim,HSCY3shear}. These simulations usually consider the shear bias as a function of several galaxies properties, e.g. signal to noise ratio (SNR), size, resolution factor, photo-z. The reliability of such calibration can be questioned in at least the following few aspects:

1. Simulations usually cannot fully capture the problems associated with instruments, neither can they faithfully replicate the setups of the image processing software;

2. There are stochastic shear biases due to, e.g., noise or PSF, that are not counted in shear-stacking/one-point statistics, but may become important for shear-shear correlations;

3. If the shear bias from certain measurement is calibrated apriori, it may be subject to additional errors caused by source selections applied posterior.   

Given these facts, it is therefore ideal to have an onsite calibration method, i.e., the shear biases are directly calibrated with the source galaxies that are in use. One way to do so is to use the field distortion (FD hereafter) effect, which is an extra image distortion caused by the optics of the telescope. By stacking the galaxy shear estimators according to their FD signals, one can directly estimate the linear shear biases. FD test has been successfully applied to the Canada-France-Hawaii Telescope Lensing Survey (CFHTLenS) data \citep{FQ_fd}, the Dark Energy Camera Legacy Survey (DECaLS) data \citep{FQ_decals}, and the HSC data \citep{FQ_HSC}. 
More recently, \cite{FQ_TC} proposed a new method called Tele-Correlation (TC hereafter). It extends the idea of FD test to the two-point statistics by cross-correlating the shear estimators of remote galaxy pairs with the same FD shear correlation signals, which can help us directly estimate the shear-shear correlation biases.

In this paper, we apply the onsite shear calibration method to the HSCpdr3 shear catalog measured by the FQ shear measurement method. In \S\ref{sec:shear_cat} we briefly introduce our shear catalogs and shear measurement methods. And we perform FD test and TC to estimate the multiplicative and additive biases for different photo-z bins, SNR cuts, optical bands, and alternative forms of shear estimators. In \S\ref{sec:2pcf} we measure the tomographic two point correlation functions after bias calibration, and try to constrain the cosmological parameters. Finally, we conclude in \S\ref{sec:conclusion}.

\section{Methodology \& Data }
\label{sec:shear_cat}

\subsection{Shear Measurement Method}

Although the proposal of this paper is suitable for shear estimators of any kind, our discussion only makes use of the shear catalog from the Fourier\_Quad (FQ hereafter) method and its alternative forms, as shown later in this section. The FQ method is a Fourier-space moment-based shear measurement algorithm. Its shear estimators are defined by the multipole moments of the galaxy power spectrum:
\begin{equation}
    \begin{aligned}
    G_{1} &=-\frac{1}{2} \int d^{2} \vec{k}\left(k_{x}^{2}-k_{y}^{2}\right) T(\vec{k}) M(\vec{k}) \\
    G_{2} &=-\int d^{2} \vec{k} k_{x} k_{y} T(\vec{k}) M(\vec{k}) \\
    N &=\int d^{2} \vec{k}\left[k^{2}-\frac{\beta^{2}}{2} k^{4}\right] T(\vec{k}) M(\vec{k}) \\
    U &=- \frac{1}{2}\beta^{2} \int d^{2} \vec{k}\left(k_{x}^{4}-6 k_{x}^{2} k_{y}^{2}+k_{y}^{4}\right) T(\vec{k}) M(\vec{k}) \\
    V &=-2 \beta^{2} \int d^{2} \vec{k}\left(k_{x}^{3} k_{y}-k_{x} k_{y}^{3}\right) T(\vec{k}) M(\vec{k})
    \end{aligned}
    \label{eq:FQ_est}
\end{equation}
in which $\vec{k}$ is the wave vector. $M(\vec{k})$ is the galaxy power spectrum considering the correction of background noise and Poisson noise. $T(\vec{k})$ converts the galaxy PSF to a Gaussian form, i.e.:
\begin{equation}
    T(\vec{k})=\left|\widetilde{W}_{\beta}(\vec{k})\right|^{2} /\left|\widetilde{W}_{P S F}(\vec{k})\right|^{2}
\end{equation}
$\widetilde{W}_{P S F}(\vec{k})$ is the power of the PSF, and $\widetilde{W}_{\beta}(\vec{k})$ is the power of the isotropic Gaussian function $W_{\beta}(\vec{x})=(2\pi\beta^2)^{-1}\exp(-|\vec{x}|^2/2\beta^2)$. The radius scale $\beta$ is set to be slightly larger than the original PSF size to avoid singularities in the conversion. More details can be found in \cite{FQ_shear}. It can be shown that the ensemble average of these estimators can achieve second order accuracy in shear recovery:
\begin{equation}
    \frac{\left\langle G_{1}\right\rangle}{\langle N\rangle}=g_{1}+O\left(g_{1,2}^{3}\right), \quad \frac{\left\langle G_{2}\right\rangle}{\langle N\rangle}=g_{2}+O\left(g_{1,2}^{3}\right)
    \label{eq:ave}
\end{equation}
Note that the averages are taken for $G_{1,2}$ and $N$ separately. 

The FQ shear estimators are proportional to the square of the galaxy flux, making it far from optimal to directly take their averages. As there are much more faint sources in the survey, the distribution of shear estimators has a high peak and a long tail, leading to a large variance when taking assemble averages. \cite{FQ_pdf} proposed another way to measure shear statistics (called PDF-SYM method). The idea is to symmetrize the following probability distribution functions (PDF):
\begin{eqnarray}
&&    \hat{G}_1=G_1- \hat{g}_1(N+U) \\ \nonumber
&&    \hat{G}_2=G_2- \hat{g}_2(N-U)
\end{eqnarray}
When the distributions of $\hat{G}_{1,2}$ are best symmetrized with respect to zero, $\hat{g}_{1,2}$ becomes a good estimate of the true shear signal $g_{1,2}$. The quantity V is kept for the transformation of U under coordinate rotation. It is proved that the PDF-SYM method can automatically reach minimum statistical error (the Cramer-Rao Bound) in shear recovery. Similarly the shear-shear correlations can also be measured by symmetrizing the joint PDF of $(\hat{G},\hat{G}^\prime)$. The PDF-SYM method has been used in various shear statistics, such as shear-shear correlation \citep{FQ_pdf,FQ_TC}, galaxy-galaxy lensing \citep{Jiaqi_ggl,matt_ggl,pedro_ggl}, shear field reconstruction \citep{Haoran_pdf}. 

As the theme of this paper is about calibration, we shall consider other forms of shear estimators for the generality of our conclusions. For convenience, we consider two alternative forms of the shear estimators based on the moments defined in FQ. They (labeled as G/N, $\mathrm{Arg(G+Ni)}$ hereafter) are defined in \cite{2026arXiv260600553L} as:

\begin{equation}
\label{eI}
e_{1,2}=\mathrm{G_{1,2}/N}, w=\frac{\mathrm{sign}(N)}{\sqrt{e_1^2+e_2^2}}
\end{equation}
and 
\begin{equation}
\label{eII}
e_{1,2}=\mathrm{Arg(G_{1,2}+Ni)}, w=\frac{1}{\sqrt{e_1^2+e_2^2}}
\end{equation}
where $\mathrm{sign}(N)$ is the sign function of N, and $\mathrm{Arg(G_{1,2}+Ni)}$ is the argument of the complex number $G_{1,2}+Ni$. 
Note that for each set, a weight $w$ is defined as in the usual shear statistics.

Note that these two new estimators have large biases, especially in the presence of noise. It is part of the purpose of this work to show that these biases can be estimated and corrected on the data directly for the shear-shear correlation using the onsite calibration, and the final results are consistent with those from the FQ shear estimators and the PDF-SYM method. The advantage of these two new estimators is that they take the forms of conventional shear estimators, i.e., the shear statistics is evaluated using the weighted sum of ellipticity-like shear estimators. They are therefore faster in shear statistics measurements than the PDF-SYM method. But as we show later in the paper, they have slightly larger statistical errors comparing to the FQ ones.

\subsection{Data}
Our shear catalog is constructed based on the wide layer imaging data from the third public data release of the Hyper Suprime-Cam Subaru Strategic Program (HSCpdr3, \cite{hscdr3}). HSCpdr3 contains 3810/3622/4625/4623/4346 exposures of images from the g/r/i/z/y band respectively, taken from Mar 2014 to Jan 2020. The sky coverage of each band is $\sim 1300 \deg^2$. The images and data products of HSCpdr3 are produced by the HSC pipeline \citep{hscpipe}. The data products are all available for downloading from the HSC official website \footnote{https://hsc-release.mtk.nao.ac.jp/}. We select the images with the prefix "CORR", which means the images have gone through a few pre-processing steps, including flatten field correction, sky background subtraction, defects detection.

The shear catalog is obtained by applying the FQ shear measurement pipeline, which evolves from the FQ shear measurement method. It involves all the steps needed for shear measurement, including background removal, astrometric calibration, source identification, PSF reconstruction, shear measurement, etc. The details of the pipeline are described in \cite{FQ_HSC}. The FQ shear catalog provides shear estimators from each individual exposure. Note that this fact allows us to record the FD information for each shear estimator, making it possible for carrying out onsite calibration when they are in use later. The final shear catalog contains shear measurements from all the five bands.

From the FD tests in \cite{FQ_HSC}, we find that the i-band shear catalog has the best performance, followed by the r \& z bands, and then the g \& y bands. We also note that using the shear catalogs from the r/i/z bands together allows us to achieve a higher SNR in shear statistics (but not much more improvement by further involving the g \& y bands). Thus, in this paper, we consider shear-shear correlation results mainly in two cases: with the i-band catalog only, or with the catalogs from the r/i/z bands together. Regarding the photo-z information, we note that the HSC team provides photo-z measured by different methods \citep{HSCphotoz}. In this work, we use the photo-z from the DEmP method \citep{demp}.

\subsection{Onsite Calibration with FD}

The FD effect causes the shape distortion of the observed object, introducing shear signals (label as FD shear, $g^f_{1,2}$, hereafter) that are typically comparable to the cosmic shear. It can be calculated by the World Coordinate System (WCS) parameters that are evaluated from astrometric calibration. If we stack the measured galaxy shear estimators (without removing the FD shear) with same FD shears, the result is supposed to be consistent with $g^f$ itself. Considering the recovered shear as a function of the FD shear, it is straightforward to estimate the bias:
\begin{equation}
    <g_{1,2}^{est}>=(1+m_{1,2})g_{1,2}^f + c_{1,2}
    \label{eq:fd_1d}
\end{equation}
where $m$ and $c$ are the multiplicative and additive biases respectively. It has been shown in \cite{FQ_fd} that the FD shear indeed provides a convenient way to test the accuracy of shear measurement and estimate the multiplicative and additive biases from the observational data directly.

The shear-shear correlation may be contaminated by the spatial correlations of the residual systematic effects on the focal plane. These spatially correlated biases cannot be detected by the original FD test. Recently, \cite{FQ_TC} proposed a method called Tele-Correlation (TC) to directly estimate the biases of shear-shear corrections. The basic idea of TC is similar to FD test, but extends to the two-point statistics. In TC, we cross-correlate the shear estimators of remote galaxy pairs which have same FD shear correlation signals. As such a correlation should not contain contributions from the astrophysical signals, the result are supposed to be consistent with $g_i^fg_i^f$. Similarly, we can write a linear bias as:
\begin{equation}
    <g_{i}^{est}g_{i}^{est}>(\Delta\theta\to\infty)= (1+\mathcal{M}_i)^2<g_{i}^fg_{i}^f> + \mathcal{C}_i
\end{equation}

Where $\mathcal{M}_i,\mathcal{C}_i$ are the multiplicative and additive biases in TC respectively. $\Delta\theta\to\infty$ means that the galaxy pairs are selected to be very far away from each other on the sky, so that their true cosmic shear correlations are negligible. For the HSCpdr3 shear catalog, we require $130^\circ\le\Delta\theta<140^\circ$. Note that in an ideal case without spatially correlated residuals, we expect $\mathcal{M}_i=m_i$ and $\mathcal{C}_i=c_i^2$, in which $m_i$ and $c_i$ are from the original FD test.

In our onsite calibration, the shear estimators are modified by the corresponding $m_{1,2}$ and $c_{1,2}$ estimated from the FD test if the purpose is to do shear stacking, or preferably the $\mathcal{M}_{1,2}$ and $\mathcal{C}_{1,2}$ from TC if it is for shear-shear correlation. In the later case, $c_{1,2}$ should be taken as the square root of $\mathcal{C}_{1,2}$. Note that if $\mathcal{C}_{1,2}<0$, $c_{1,2}$ is ill-defined, and therefore cannot be directly applied on the shear catalog. Instead, one can add $\pm\sqrt{-\mathcal{C}_{1,2}}$ to the two shear estimators of the galaxy pair respectively when measuring the shear-shear correlation.

In our final shear catalog, the FD signals have been subtracted. Thus we should add $g^f$ back to shear estimators before FD test and TC:
\begin{eqnarray}
&&    G_1^{f}=G_1+(g_1^f+c_1')(N+U)+(g_2^f+c_2')V\\ \nonumber
&&    G_2^{f}=G_2+(g_2^f+c_2')(N-U)+(g_1^f+c_1')V
\end{eqnarray}
Note that $c_{1,2}'$ are the additive biases from the calibration of the whole source sample when preparing for the shear catalog. 
After the estimation of $m,c$ for the source sample in use, these biases together with $g^f$ should be applied to the shear estimators in the following form:
\begin{eqnarray}
   G_1^{cal}&=&\frac{G_1^f-c_1(N+U)-c_2V}{1+m_1}-g_1^f(N+U)-g_2^fV \\ \nonumber
   G_2^{cal}&=&\frac{G_2^f-c_2(N-U)-c_1V}{1+m_2}-g_2^f(N-U)-g_1^fV 
\end{eqnarray}
Similarly, for the alternatively defined shear estimators ($\mathrm{G/N}$ or $\mathrm{Arg(G+Ni)}$), we should use $G_{1,2}^{f}$ to define $e_{1,2}^{f}$ according to eq.(\ref{eI}) or (\ref{eII}), and apply the calibration results as:
\begin{equation}
    e_{1,2}^{cal}=\frac{e_{1,2}^{f}-c_{1,2}}{1+m_{1,2}}-g_{1,2}^f
\end{equation}
Note that the weights are not updated, and still defined using $G_{1,2}^{f}$.

We caution the readers that although shear measurements of other methods typically do not involve the above procedures, and the FD signals are believed to be removed once for all, residual shear systematic errors proportional to the FD signals at the source location should exit when the multiplicative bias due to posterior source selection shows up.

\subsection{Calibration Results}

We apply several cuts to select galaxies for correlation measurements. The cuts are summarized as follows: We choose galaxies with SNR larger than 10. For the purpose of consistency tests, higher SNR cuts (15, 20) are also applied. In addition, we exclude galaxies with large FD signals: $\sqrt{(g^f_1)^2+(g^f_2)^2}>0.02$, which are located at the edge of the exposures where the PSF reconstruction is not reliable enough \citep{FQ_HSC}. We also remove galaxies measured on problematic chips with id 33,43,94. Besides, we introduce a photo-z error cut $z_{err}\le0.05(1+z)$ to control the photo-z uncertainty.

\begin{figure}[htbp]
    \centering
    \includegraphics[width=\textwidth]{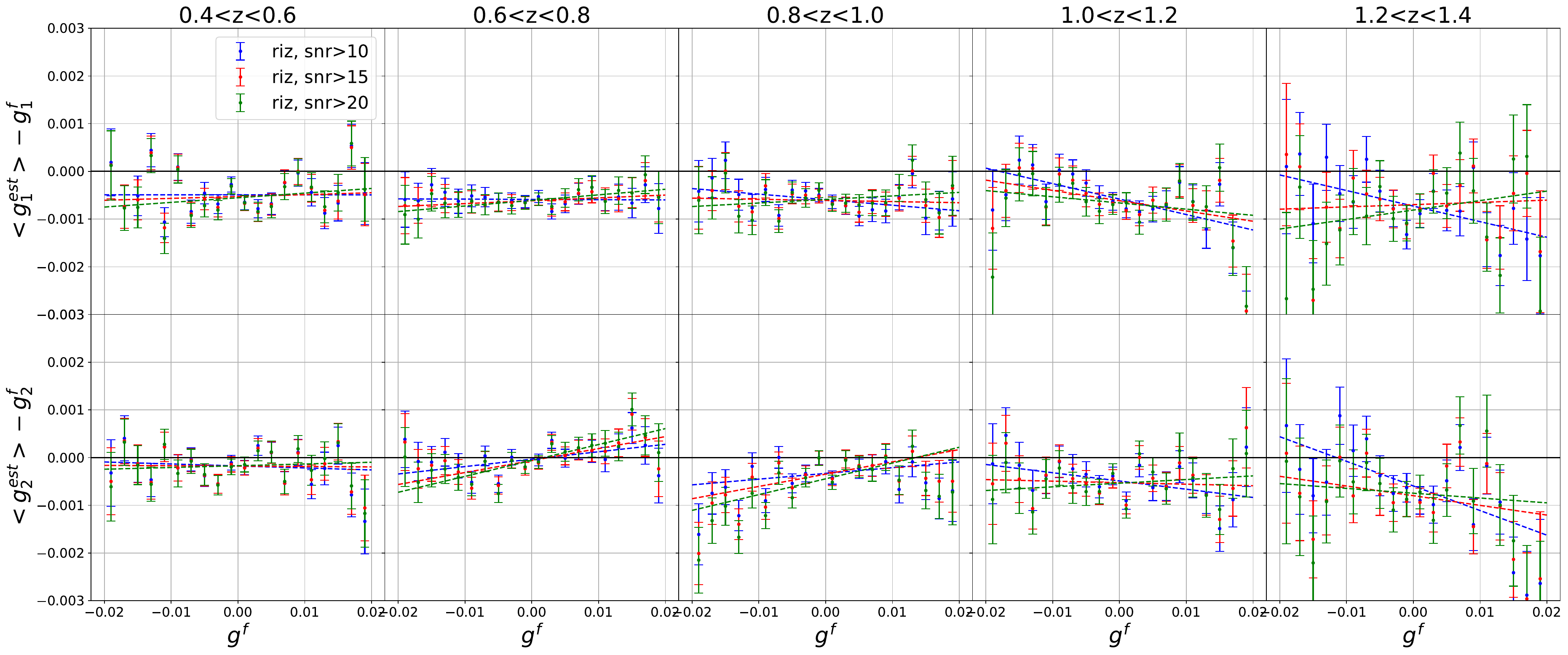}
    \caption{The results of the r/i/z bands FD tests for different redshift bins and shear components ($g_1$,$g_2$) using FQ estimators. In each panel, the data points represent the stacked shear signals from galaxies binned according to their FD shear values ($g^f_1$ or $g^f_2$). The dashed lines show the best-fit results. Different colors represent different SNR cut. To clearly show the data points and the best-fit lines, we plot $<g_i^{est}>-g_i^f$ against $g_i^f$ ($i=1,2$) in this figure.}
    \label{fig:fd_1d_riz_pdf}
\end{figure}

For the purpose of tomographic cosmic shear analysis, we divide the galaxies into 5 photo-z bins: $[0.4,0.6]$, $[0.6,0.8]$, $[0.8,1.0]$, $[1.0,1.2]$, $[1.2,1.4]$. First, we apply FD test for the r/i/z bands shear catalog under different SNR cuts and photo-z bins using FQ estimators. The results are shown in Fig.\ref{fig:fd_1d_riz_pdf} and the estimated shear biases are summarized in Tab.\ref{tab:fd_1d_riz_mc}. As shown in the figure and table, the estimated biases vary under different SNR and photo-z cut, which indicates that shear measurement biases depend on galaxy selections. Thus it is necessary to evaluate the shear biases for each galaxy sample in use.

\begin{table}[htb!]
    \centering
    \begin{tabular}{cccccc}
        \hline
        \hline
        photo-z&SNR cut&\quad$10^2m_1$\quad&\quad$10^4c_1$\quad&\quad$10^2m_2$\quad&\quad$10^4c_2$ \\
        \hline
           &$>10$&$-0.0\pm0.7$&$-4.9\pm0.6$&$-0.4\pm0.7$&$-1.8\pm0.6$ \\
$[0.4-0.6]$&$>15$&$0.4\pm0.7$&$-5.3\pm0.6$&$-0.1\pm0.8$&$-1.8\pm0.6$ \\
           &$>20$&$1.0\pm0.8$&$-5.5\pm0.6$&$0.4\pm0.8$&$-1.7\pm0.6$ \\ \hline
           &$>10$&$-0.1\pm0.6$&$-5.9\pm0.5$&$1.6\pm0.6$&$-0.3\pm0.5$ \\
$[0.6-0.8]$&$>15$&$0.6\pm0.6$&$-6.2\pm0.5$&$2.5\pm0.6$&$-0.6\pm0.5$ \\
           &$>20$&$1.2\pm0.6$&$-6.1\pm0.5$&$3.3\pm0.7$&$-0.6\pm0.5$ \\ \hline
           &$>10$&$-1.2\pm0.7$&$-5.9\pm0.5$&$1.2\pm0.7$&$-3.3\pm0.5$ \\
$[0.8-1.0]$&$>15$&$-0.2\pm0.7$&$-6.1\pm0.5$&$2.6\pm0.7$&$-3.5\pm0.5$ \\
           &$>20$&$0.7\pm0.7$&$-5.9\pm0.6$&$3.3\pm0.7$&$-4.5\pm0.6$ \\ \hline
           &$>10$&$-3.2\pm0.9$&$-5.8\pm0.7$&$-1.7\pm0.9$&$-4.9\pm0.7$ \\
$[1.0-1.2]$&$>15$&$-2.2\pm0.9$&$-6.1\pm0.7$&$-0.3\pm0.9$&$-5.3\pm0.7$ \\
           &$>20$&$-1.3\pm1.0$&$-6.6\pm0.8$&$0.8\pm1.0$&$-5.4\pm0.8$ \\ \hline
           &$>10$&$-3.3\pm1.4$&$-7.3\pm1.1$&$-5.2\pm1.5$&$-5.9\pm1.1$ \\
$[1.2-1.4]$&$>15$&$0.5\pm1.5$&$-7.0\pm1.2$&$-2.0\pm1.5$&$-8.0\pm1.2$ \\
           &$>20$&$2.0\pm1.8$&$-8.1\pm1.4$&$-1.0\pm1.9$&$-7.5\pm1.4$ \\ \hline
    \end{tabular}
    \caption{The estimated multiplicative and additive biases from the r/i/z bands FD tests under different SNR and photo-z cuts using FQ estimators.}
    \label{tab:fd_1d_riz_mc}
\end{table}

\begin{figure}[htbp]
    \centering
    \includegraphics[width=\textwidth]{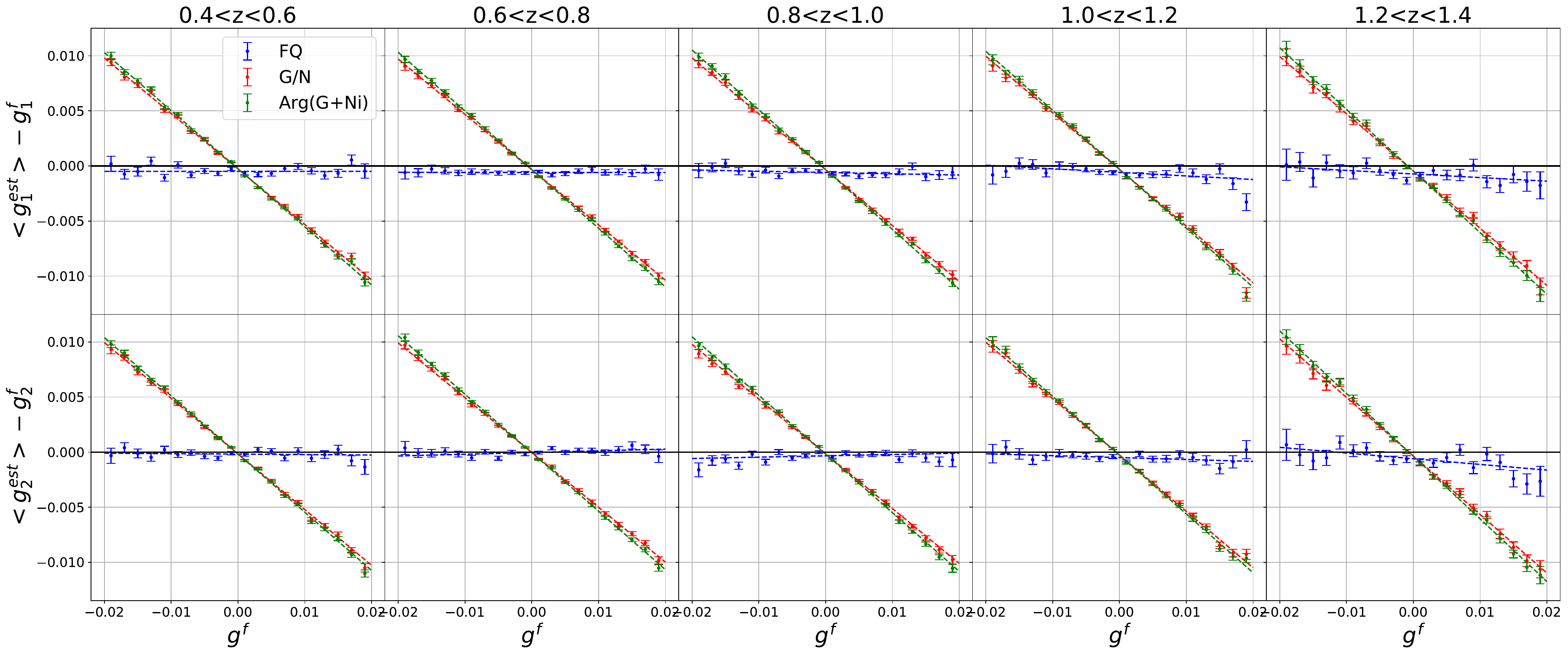}
    \caption{Similar to Fig.\ref{fig:fd_1d_riz_pdf}, but for different shear estimators with $\mathrm{SNR}>10$ cut. Different colors represent different shear estimators in this figure.}
    \label{fig:fd_1d_i_gn}
\end{figure}

For comparison, we also apply the FD test for other shear estimators. In this test, we fix $\mathrm{SNR}>10$ cut, and introduce resutls from diffrent shear estimators (FQ,$\mathrm{G/N}$ and $\mathrm{Arg(G+Ni)}$). The results are shown in Fig.\ref{fig:fd_1d_i_gn} and Tab.\ref{tab:fd_2d_i_mc}. We find that compared to the FQ results, the $\mathrm{G/N}$ and $\mathrm{Arg(G+Ni)}$ shear estimators generally have larger bias values.

\begin{table}[htb!]
    \setlength{\tabcolsep}{2pt}
    \centering
    \resizebox{\textwidth}{!}{
    \begin{tabular}{cccccccccc}
        \hline
        \hline
        photo-z&estimators&$10^2m_1$&$10^4c_1$&$10^2m_2$&$10^4c_2$&$10^2\mathcal{M}_1$&$10^7\mathcal{C}_1$&$10^2\mathcal{M}_2$&$10^7\mathcal{C}_2$ \\
        \hline
           &$\mathrm{FQ}$&$-0.0\pm0.7$&$-4.9\pm0.6$&$-0.4\pm0.7$&$-1.8\pm0.6$&$-1.9\pm0.7$&$26.7\pm24.8$&$-3.5\pm0.6$&$-34.9\pm21.8$ \\
$[0.4-0.6]$&$\mathrm{G/N}$&$-50.3\pm0.4$&$-2.7\pm0.3$&$-50.5\pm0.4$&$-1.6\pm0.3$&$-35.9\pm0.1$&$-0.4\pm4.5$&$-36.5\pm0.2$&$0.4\pm5.1$ \\
           &$\mathrm{Arg(G+Ni)}$&$-52.7\pm0.4$&$-2.5\pm0.3$&$-52.8\pm0.4$&$-1.7\pm0.3$&$-37.4\pm0.1$&$-0.4\pm3.9$&$-37.9\pm0.1$&$0.4\pm4.5$ \\ \hline
           &$\mathrm{FQ}$&$-0.1\pm0.6$&$-5.9\pm0.5$&$1.6\pm0.6$&$-0.3\pm0.5$&$-2.1\pm0.6$&$43.2\pm20.3$&$-1.7\pm0.6$&$-36.2\pm21.3$ \\
$[0.6-0.8]$&$\mathrm{G/N}$&$-50.2\pm0.3$&$-3.4\pm0.3$&$-49.7\pm0.4$&$-0.3\pm0.3$&$-36.1\pm0.1$&$-0.4\pm4.4$&$-36.3\pm0.1$&$0.2\pm4.6$ \\
           &$\mathrm{Arg(G+Ni)}$&$-53.2\pm0.3$&$-3.3\pm0.2$&$-53.1\pm0.3$&$-0.5\pm0.2$&$-38.0\pm0.1$&$-0.3\pm3.8$&$-38.3\pm0.1$&$0.2\pm4.0$ \\ \hline
           &$\mathrm{FQ}$&$-1.2\pm0.7$&$-5.9\pm0.5$&$1.2\pm0.7$&$-3.3\pm0.5$&$-4.8\pm0.7$&$9.4\pm22.9$&$2.2\pm0.7$&$-99.6\pm25.5$ \\
$[0.8-1.0]$&$\mathrm{G/N}$&$-50.7\pm0.4$&$-3.4\pm0.3$&$-49.7\pm0.4$&$-1.5\pm0.3$&$-36.8\pm0.1$&$-0.2\pm4.9$&$-36.1\pm0.1$&$2.3\pm3.9$ \\
           &$\mathrm{Arg(G+Ni)}$&$-54.3\pm0.4$&$-3.3\pm0.3$&$-53.2\pm0.4$&$-1.7\pm0.3$&$-38.8\pm0.1$&$-0.2\pm4.1$&$-38.3\pm0.1$&$2.1\pm3.4$ \\ \hline
           &$\mathrm{FQ}$&$-3.2\pm0.9$&$-5.8\pm0.7$&$-1.7\pm0.9$&$-4.9\pm0.7$&$-8.8\pm0.8$&$35.2\pm26.8$&$-1.8\pm0.9$&$-58.5\pm27.9$ \\
$[1.0-1.2]$&$\mathrm{G/N}$&$-51.4\pm0.5$&$-3.2\pm0.4$&$-51.1\pm0.5$&$-2.6\pm0.4$&$-38.1\pm0.2$&$-1.9\pm6.1$&$-37.3\pm0.2$&$1.3\pm6.5$ \\
           &$\mathrm{Arg(G+Ni)}$&$-53.5\pm0.5$&$-3.0\pm0.4$&$-53.3\pm0.5$&$-2.7\pm0.4$&$-39.6\pm0.2$&$-1.5\pm5.4$&$-38.7\pm0.2$&$0.9\pm5.6$ \\ \hline
           &$\mathrm{FQ}$&$-3.3\pm1.4$&$-7.3\pm1.1$&$-5.2\pm1.5$&$-5.9\pm1.1$&$-7.7\pm1.2$&$23.9\pm36.7$&$-5.1\pm1.3$&$-17.6\pm43.4$ \\
$[1.2-1.4]$&$\mathrm{G/N}$&$-51.9\pm0.8$&$-4.6\pm0.6$&$-53.1\pm0.8$&$-3.4\pm0.6$&$-37.5\pm0.3$&$-0.2\pm11.7$&$-38.7\pm0.3$&$0.7\pm8.9$ \\
           &$\mathrm{Arg(G+Ni)}$&$-55.9\pm0.7$&$-4.6\pm0.6$&$-56.9\pm0.7$&$-3.7\pm0.6$&$-40.2\pm0.3$&$0.4\pm9.2$&$-40.9\pm0.2$&$1.2\pm7.1$ \\ \hline
    \end{tabular}}
    \caption{The summarized m,c estimated by FD test and $\mathcal{M,C}$ estimated by TC from the r/i/z bands shear catalog under $\mathrm{SNR}>10$ cut.}
    \label{tab:fd_2d_i_mc}
\end{table}

\begin{figure}[htbp]
    \centering
    \includegraphics[width=\textwidth]{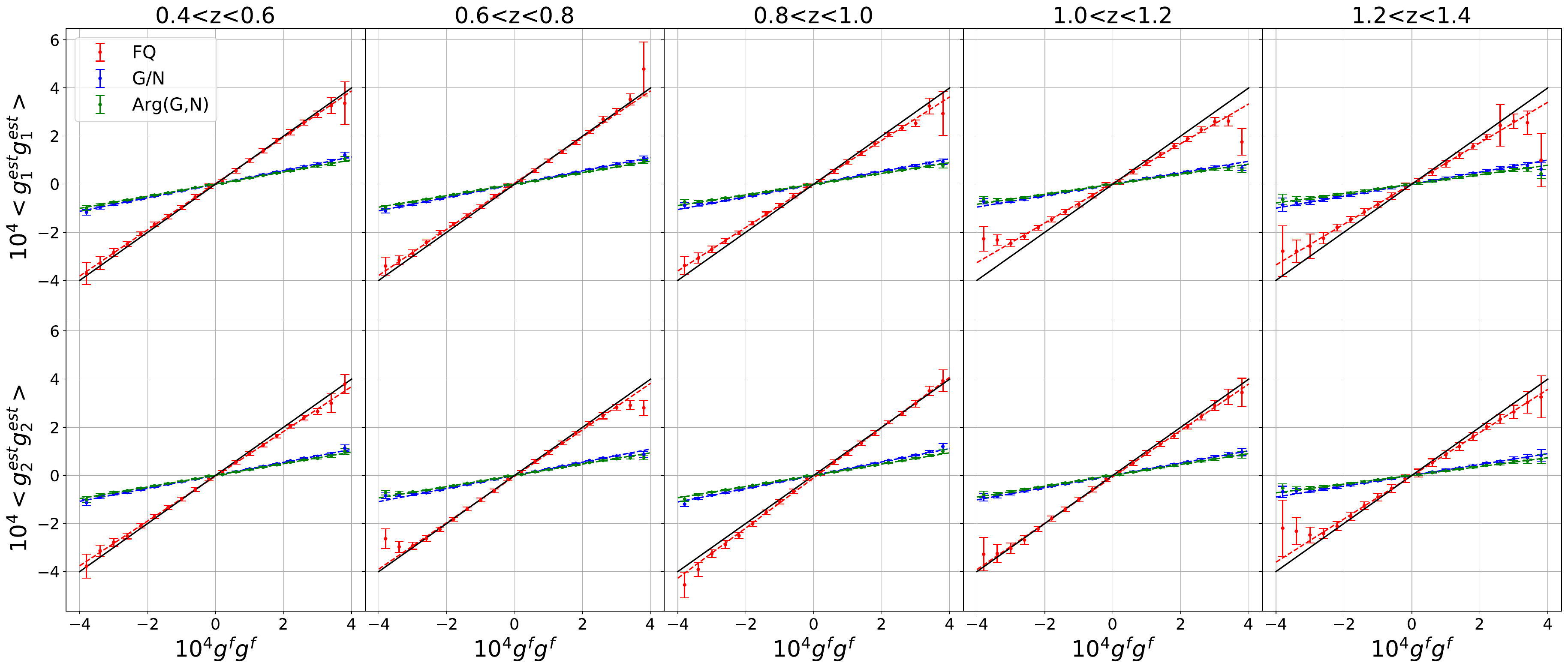}
    \caption{The TC results of the r/i/z bands shear catalog under $\mathrm{SNR}>10$ cut. Each column represents different redshift bins, and the two rows represent two shear components.  In each panel, the data points represent the TC signals. The dashed lines show the best-fit results, and black line refers to $y=x$. Different colors represent different shear estimators.}
    \label{fig:fd_2d_i}
\end{figure}

Finally, we use TC to estimate the shear bias for the correlation measurement. Here we only show the results with $\mathrm{SNR}>10$ cut in Fig.\ref{fig:fd_2d_i} and Tab.\ref{tab:fd_2d_i_mc}, as the other selection criteria yield similar results. From the figure and table, we find that the $\mathcal{M}$ also vary in different photo-z bins for all three shear estimators, indicating that selection effects also affect the results of TC. Similarly to the results from the FD test, we find significant $\mathcal{M}$ for the shear estimators of $\mathrm{G/N}$ and $\mathrm{Arg(G+Ni)}$. For the additive bias, the $\mathcal{C}$ values are large, but the corresponding errorbars are also large especially comparing to the errorbars of $c^2$. Meanwhile we test the TC results with other angular separation of galaxy pairs, the estimated $\mathcal{M}$ is quite stable, but $\mathcal{C}$ varies in $~1-2\sigma$. Thus we conclude that TC does not constrain $\mathcal{C}$ very well, and we will not use $\mathcal{C}$ to calibrate the correlation results. Compared to Fig.\ref{fig:fd_1d_riz_pdf}, we can see that for the FQ method, the estimated $\mathcal{M}$ are generally consistent with the $m$ values from FD test, but for $\mathrm{G/N}$ and $\mathrm{Arg(G+Ni)}$, $\mathcal{M}$ and $m$ can be significantly different (more than $10\sigma$). It indicates that the influence of correlated systematic biases varies for different estimators. Thus we decide to use $m,c$ from FD test to calibrate shear-shear correlations for the FQ method, and use $\mathcal{M}$ from TC and c from FD test for the $\mathrm{G/N}$ and $\mathrm{Arg(G+Ni)}$ estimators.

\section{shear shear correlation}
\label{sec:2pcf}
\subsection{2PCF}
The 2-point shear-shear correlation function (2PCF) is defined as:
\begin{equation}
    \xi_{ \pm}(\Delta\theta)=\left\langle\gamma_{t}(\mathbf{\theta}) \gamma_{t}(\mathbf{\theta}+\Delta \mathbf{\theta})\right\rangle \pm\left\langle\gamma_{\times}(\mathbf{\theta}) \gamma_{\times}(\mathbf{\theta}+\Delta \mathbf{\theta})\right\rangle
\end{equation}
where $\gamma_t$ and $\gamma_{\times}$ are the tangential and cross components of the shear respectively, with respect to the direction connecting the two galaxies in a pair. Given that the FQ shear estimators are measured on individual exposures, we only use the shear estimators from two different exposures to measure their correlations. This is for the purpose of avoiding possible correlated systematic errors on the exposures.

As mentioned in Section \ref{sec:shear_cat}, we divide the shear catalogs into 5 photo-z bins for tomography measurement. We calculate both auto-correlations and cross-correlations among the five photo-z bins, resulting in 15 correlation functions. These correlations are measured in eight logarithmic bins of angular distance ranging from 3 to 108 arcmin. The covariance matrix is estimated by jackknife with 200 subsamples. The small scale cut is determined in order to control the baryonic effects and the large scale cut is determined by the size of jackknife samples.

We adopt the onsite calibration method to calibrate the results of 2PCFs. Referring to Tab \ref{tab:fd_1d_riz_mc} we find that most redshift bins tend to have negative m and non-zero c, thus the calibration of $\xi_+$ behaves as an increase on small scales and a suppression on large scales. As $\xi_-$ is almost not affected by c, the calibration behaves as an increase on all scales. And because m,c depend on the redshift selections, the onsite calibration varies in different redshift combinations. 

\begin{figure}[htbp]
    \centering
    \includegraphics[width=\textwidth]{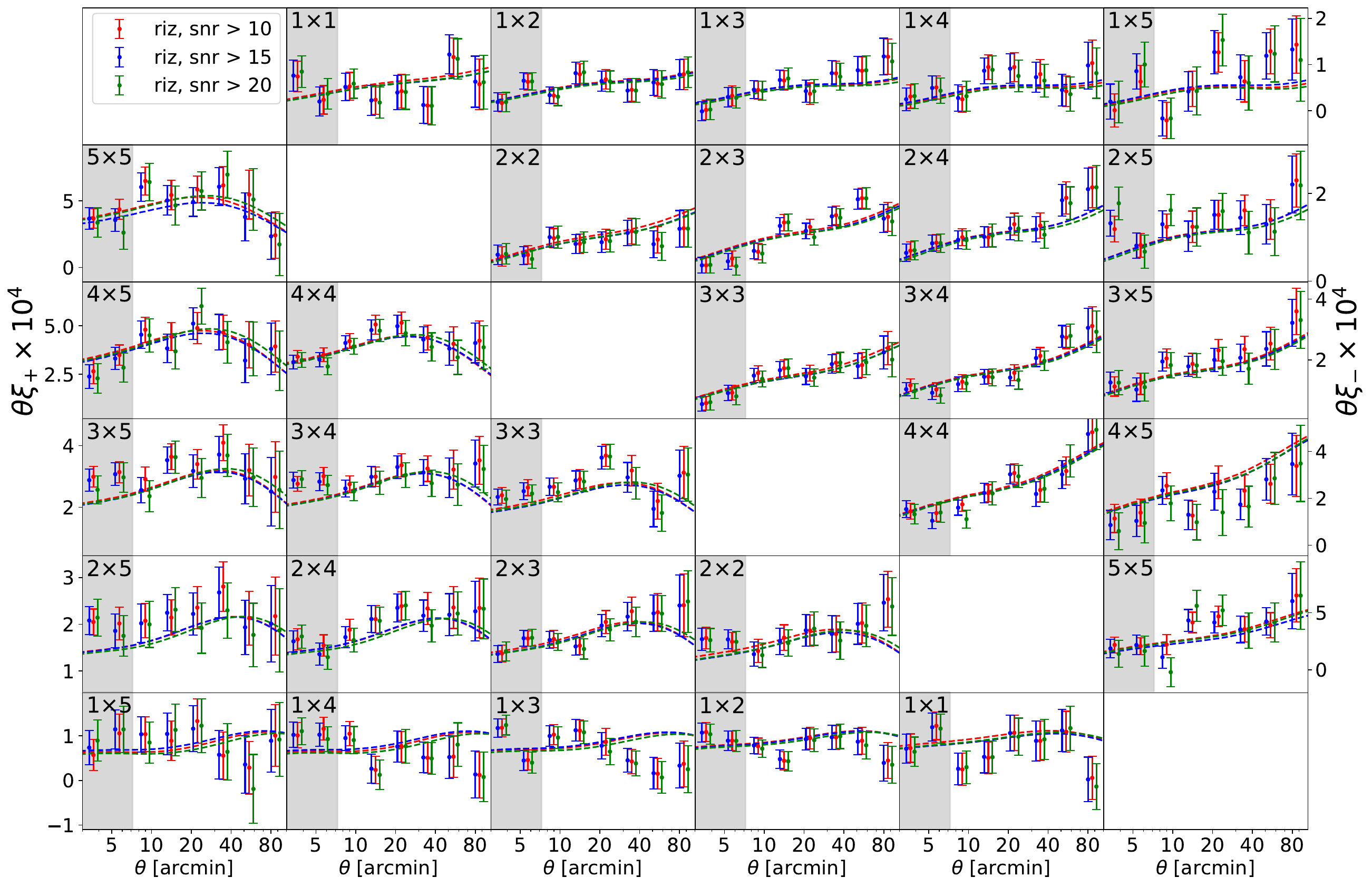}
    \caption{The 2PCFs for the FQ estimators after onsite calibration, using the shear catalogs of the r/i/z bands. Each panel represents different redshift bin (labelled as 1-5) combinations, and the left-bottom and top-right panels represent $\xi_+$ and $\xi_-$ respectively. Different colors represent different SNR cut. The error bars are estimated by the jackknife method. The dashed lines show the corresponding best-fit model. The unshaded region refers to the fiducial scale cut.}
    \label{fig:2pcf_riz}
\end{figure}

In Fig. \ref{fig:2pcf_riz}, We plot the final calibrated 2PCFs with r/i/z bands shear catalog and various SNR cuts using FQ estimators. As shown in this figure, the results from different SNR cuts are generally consistent with each other. The total SNR of 2PCFs with $\mathrm{SNR}>15,20$ cut is $\sim5,15\%$ smaller than $\mathrm{SNR}>10$ cut, which is expected because more galaxies are involved with lower SNR cut. We also tried SNR cuts lower than 10, but find that the errorbars do not reduce significantly, indicating that systematic errors become dominant. Thus, we decide to use 10 as our lowest SNR cut.

\begin{figure}[htbp]
    \centering
    \includegraphics[width=\textwidth]{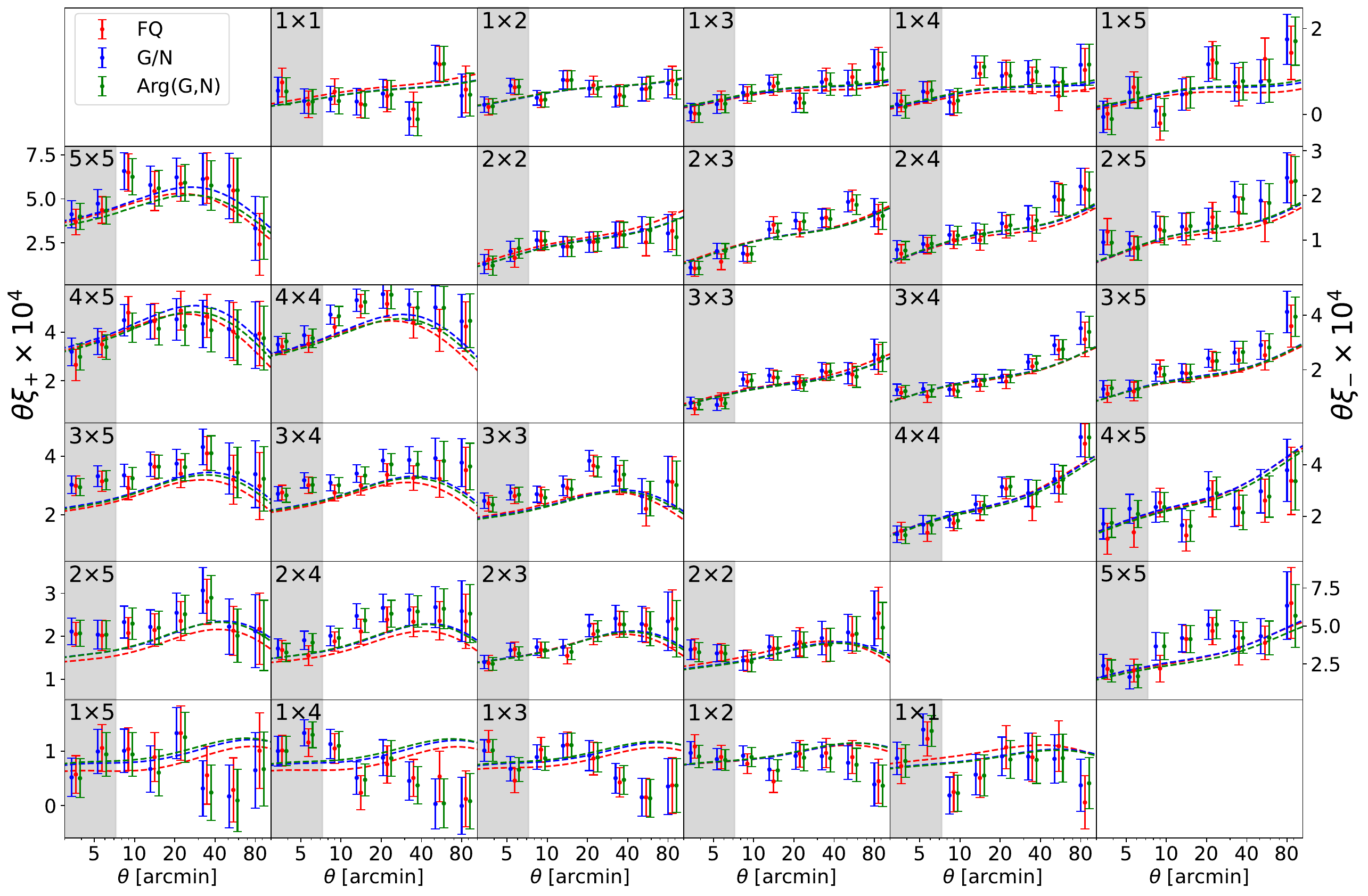}
    \caption{Similar to Fig.\ref{fig:2pcf_riz}, but for different shear estimators, all with $\mathrm{SNR}>10$ cut.}
    \label{fig:2pcf_gn}
\end{figure}

we also test the consistency of 2PCFs under various shear estimators. The results are shown in Fig. \ref{fig:2pcf_gn}. In this Figure, we fix the $\mathrm{SNR}>10$ cut, the other SNR selections have similar properties. As shown in this figure, the $\mathrm{G/N}$ and $\mathrm{Arg(G+Ni)}$ results are generally consistent with those of FQ.

\subsection{B mode test}
\label{Bmode}

As the shear field is expected to be curl free, the B mode test provides a useful way to check the residual systematics in shear measurement. Although in 2PCF, there might be physical B modes from second order lensing deflection, intrinsic alignment, redshift clustering. These effects are expected to be oder of magnitude lower than E mode. Following \cite{2pcf_Bmode}, we separate the E mode and the B mode components as:

\begin{equation}
    \begin{aligned}
    \xi_{+}^E(\theta) & =\frac{1}{2}\left[\xi_{+}(\theta)+\xi_{-}(\theta)+\int_\theta^{\infty} \frac{\mathrm{d} \phi}{\phi} \xi_{-}(\phi)\left(4-12 \frac{\theta^2}{\phi^2}\right)\right], \\
    \xi_{-}^E(\theta) & =\frac{1}{2}\left[\xi_{+}(\theta)+\xi_{-}(\theta)+\int_0^\theta \frac{\mathrm{d} \phi \phi}{\theta^2} \xi_{+}(\phi)\left(4-12 \frac{\phi^2}{\theta^2}\right)\right], \\
    \xi_{+}^B(\theta) & =\frac{1}{2}\left[\xi_{+}(\theta)-\xi_{-}(\theta)-\int_\theta^{\infty} \frac{\mathrm{d} \phi}{\phi} \xi_{-}(\phi)\left(4-12 \frac{\theta^2}{\phi^2}\right)\right], \\
    \xi_{-}^B(\theta) & =\frac{1}{2}\left[\xi_{+}(\theta)-\xi_{-}(\theta)+\int_0^\theta \frac{\mathrm{d} \phi \phi}{\theta^2} \xi_{+}(\phi)\left(4-12 \frac{\phi^2}{\theta^2}\right)\right],
    \end{aligned}
\end{equation}

To calculate these integrals, we separate them into two parts. For $\theta$ from 3 arcmins to 108 arcmins, we measure the correlation functions in a much narrow (64) logarithmic bins and take a Riemann sum. For theta lower than 3 and higher than 108 arcmin, we use the best fit model from cosmology constraint. And the errorbars are estimated with jackknife.

\begin{figure}[htbp]
    \centering
    \includegraphics[width=\textwidth]{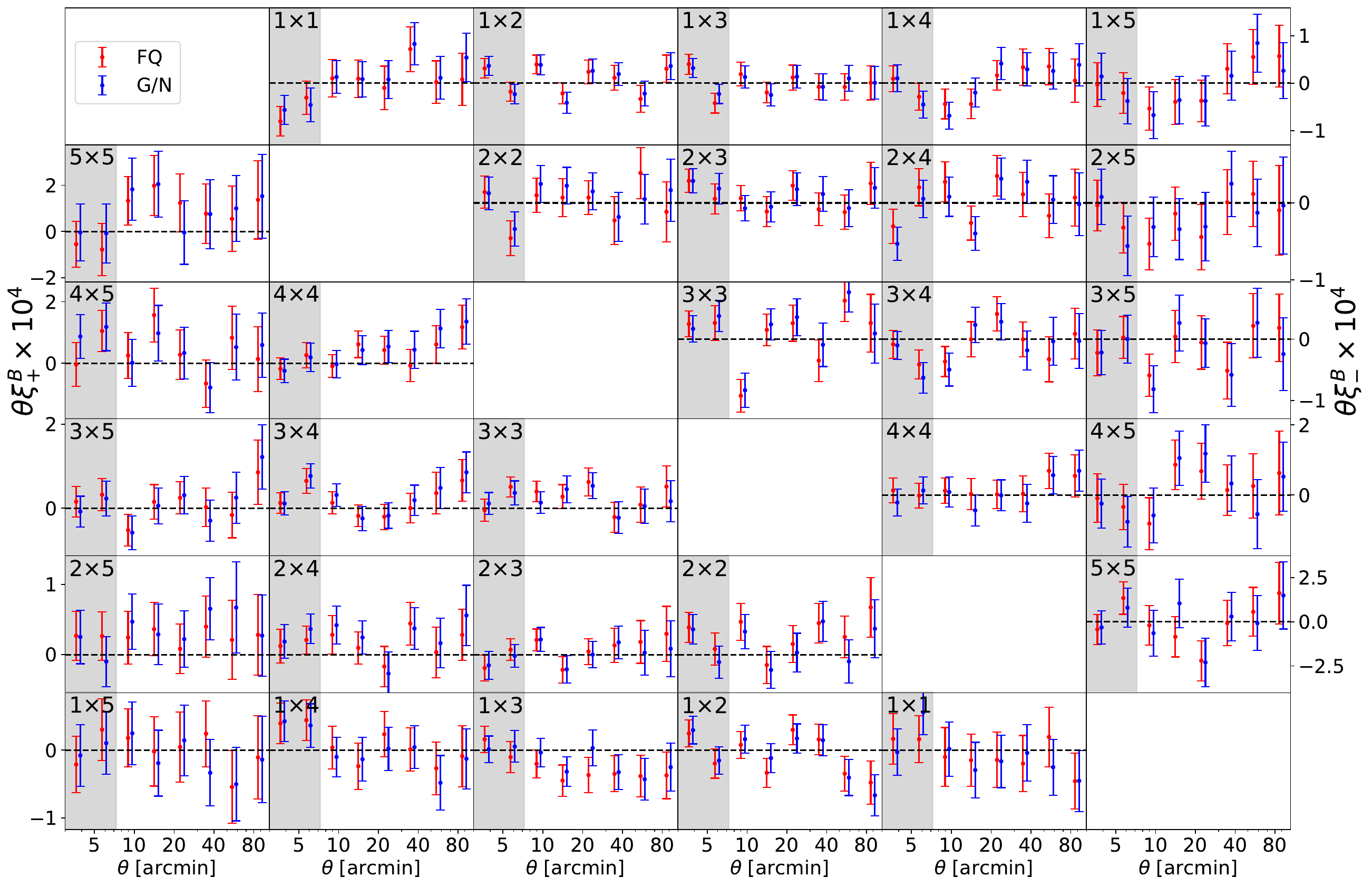}
    \caption{The B mode of 2PCFs for FQ and G/N estimators with r/i/z bands catalogs and $\mathrm{SNR}>10$ cut. The errorbars are estimated with jackknife. The unshaded region refers to the fiducial scale cut.}
    \label{fig:bmode_i}
\end{figure}

Here we show the B mode results of FQ and $\mathrm{G/N}$ estimators from the r/i/z bands shear catalog with $\mathrm{SNR}>10$ cut as an example in Fig.\ref{fig:bmode_i}. The other selections have similar properties. As shown in this figure, the B mode signals are generally consistent with zero within errorbars for all redshift bin combinations, indicating that the residual systematics are well controlled in our 2PCF measurements. Note that in these results, the shear biases from the FD test or TC have been applied.

\subsection{cosmological constraints}
\label{mcmc}

With flat sky approximation, the shear shear correlation functions are related to the angular power spectrum via Hankel transform:
\begin{equation}
    \xi_{ \pm}^{i j}(\theta) =\int_0^{\infty} \frac{d \ell}{2 \pi} \ell C^{i j}(\ell) J_{0/4}(\ell \theta)
\end{equation}
where $J_{0/4}$ is the 0th/4th order Bessel function of the first kind. The observed galaxy shapes are determined by both the gravitational lensing effect and the intrinsic alignments (IA, \cite{2004PhRvD..70f3526H,2015PhR...558....1T}) of galaxies. Thus the angular power spectrum contains contributions from both lensing and IA:

\begin{equation}
    C^{ij}(\ell)=C_{GG}^{ij}(\ell)+C_{GI}^{ij}(\ell)+C_{GI}^{ji}(\ell)+C_{II}^{ij}(\ell)
\end{equation}
where the $C_{GG}^{ij}(\ell)$ and $C_{II}^{ij}(\ell)$ is the lensing-lensing and IA-IA auto power spectra, and the $C_{GI}^{ij}(\ell)$ are the lensing-IA cross power spectra.

In a spatially flat universe, the lensing angular power spectra can be related to the matter power spectrum via the Limber approximation:
\begin{equation}
    C_{GG}^{i j}(\ell) =\int_0^{\infty} d z \frac{q^i(z) q^j(z)}{\chi(z)^2} P_\delta\left(\frac{\ell}{\chi(z)}, z\right)
\end{equation}
where $\chi(z)$ is the comoving distance to redshift $z$, $P_\delta(k,z)$ is the matter power spectrum at wave number $k$ and redshift $z$. The lensing kernel $q^i(z)$ for the $i^{th}$ redshift bin is defined as:
\begin{equation}
    q_i(\chi)=\frac{3}{2} \Omega_{\mathrm{m}}\left(\frac{H_0}{c}\right)^2 \frac{\chi}{a(\chi)} \int_\chi^{\chi_H} d \chi^{\prime} n_i\left(\chi^{\prime}\right) \frac{\chi^{\prime}-\chi}{\chi^{\prime}},
\end{equation}
where $\Omega_m$ and $H_0$ are the matter density and Hubble parameter at redshift zero, a is the scale factor, c is the light speed, and $n_i(\chi)$ is the normalized redshift distribution of the galaxy in the $i^{th}$ redshift bin.

\begin{figure}[htbp]
    \centering
    \includegraphics[width=0.6\textwidth]{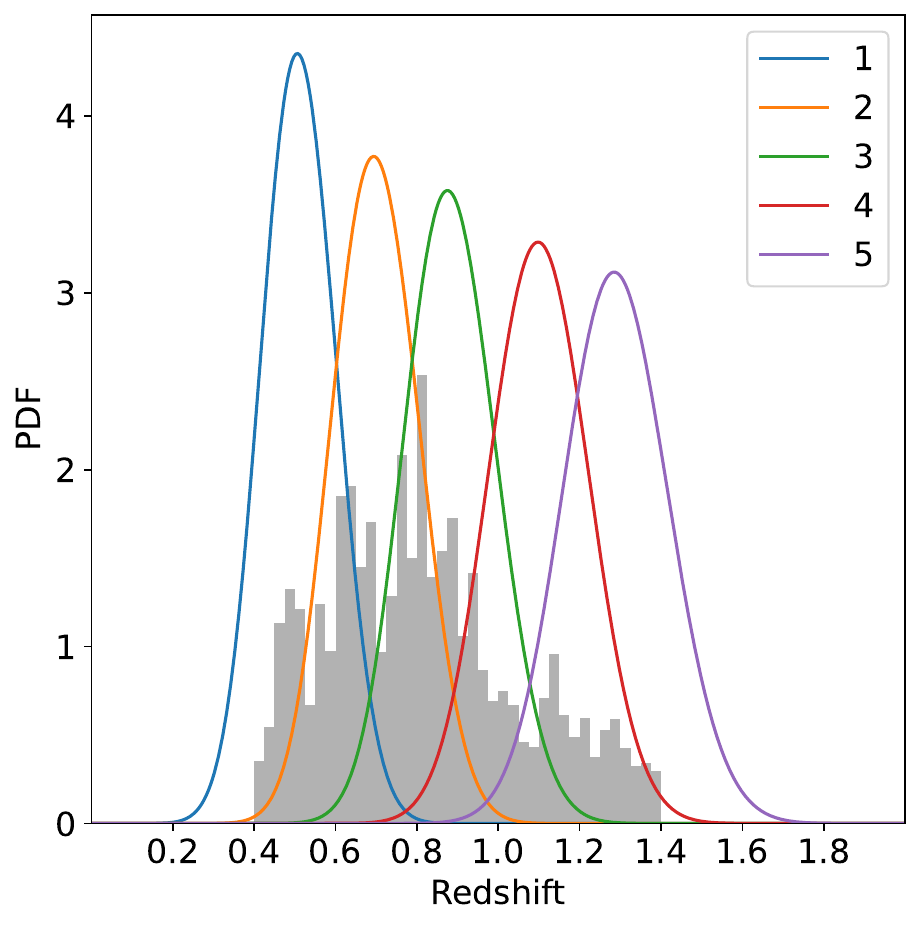}
    \caption{The normalized redshift distributions of five photo-z bins. Different colors represent different redshift bins. The gray histogram shows the overall photo-z distribution.}
    \label{fig:nz}
\end{figure}

We use the photo-z measured with DEMP method \citep{demp}, which are provided by the official data release \citep{HSCphotoz}. We cut galaxies whose photo-z error $z_{err}$ are less than $0.05(1+z)$ to exclude galaxies with large photo-z uncertainty. We compute the true redshift distribution by simply convolving the photo-z ($z_p$) distribution with a Gaussian kernel:
\begin{equation}
    n(z)=\int dz_p n(z_p)\frac{1}{\sqrt{2\pi}\sigma_z}\exp\left(-\frac{(z-z_p)^2}{2\sigma_z^2}\right)
\end{equation}
where $\sigma_z=0.05(1+z_p)$, the factor 0.05 is set according to \cite{HSCphotoz}. The final normalized redshift distributions of r/i/z bands shear catalog with cut $\mathrm{SNR}>10$ in five photo-z bins are shown in Fig.\ref{fig:nz}.

Similarly, the IA related angular power spectra can be expressed as:
\begin{equation}
    \begin{aligned}
        C_{GI}^{i j}(\ell) &=\int_0^{\infty} d z \frac{q^i(z) n^j(z)}{\chi(z)^2} P_{GI}\left(\frac{\ell}{\chi(z)}, z\right) \\
        C_{II}^{i j}(\ell) &=\int_0^{\infty} d z \frac{n^i(z) n^j(z)}{\chi(z)^2} P_{II}\left(\frac{\ell}{\chi(z)}, z\right)
    \end{aligned}
\end{equation}
In this paper, we adopt the Nonlinear Alignment model (NLA, \cite{NLA}) to model the II and GI power spectrum:
\begin{equation}
    \begin{aligned}
        P_{GI}(k, z) & =c^2(z) P_{\delta}(k, z) \\
        P_{II}(k, z) & =c(z) P_{\delta}(k, z)
    \end{aligned}
\end{equation}
where $c(z)$ is given by:
\begin{equation}
    c(z)=-A_{\mathrm{IA}} \frac{C_1 \rho_{\mathrm{crit}}\Omega_{\mathrm{m}}}{D(z)}
\end{equation}
where $A_{\mathrm{IA}}$ is a free parameter to describe the amplitude of IA. $\rho_{crit}$ is the critical density, $D(z)$ is the growth factor. $C_1$ is a normalization constant, which is set to $5 \times 10^{-14}h^{-2}M_\odot^{-1} Mpc^3$ to match the observational results \citep{NLA_C1}.

For the $P_\delta(k,z)$, we use the nonlinear matter power spectrum from the halofit model \citep{halofit_1,halofit_2}. The overall calculation is performed with the Core Cosmology Library (pyCCL, \footnote{https://github.com/LSSTDESC/CCL} \cite{pyccl}).
We use the Markov Chain Monte Carlo(MCMC) method to sample the posterior distribution of cosmological and systematic parameters. The likelihood is assumed to be Gaussian.

As reported in \cite{Hartlap_factor}. the inverse of the covariance matrix estimated from jackknife method is biased. Thus we apply the Hartlap correction to the inverse covariance matrix:
\begin{equation}
    C^{-1}=\frac{N_{jk}-N_{data}-2}{N_{jk}-1}C_{jk}^{-1}
\end{equation}
where $N_{jk}$ is the number of jackknife subsamples, and $N_{data}$ is the number of data points. This correction acquires $N_{data}<N_{jk}-2$. Thus we throw away two lowest angular bins for each 2PCF, resulting in 180 data points in total. Meanwhile, we find that the $C_{jk}$ is noisy due to limited number of jackknife subsamples. Thus we take the pseudo inverse \citep{moore1920reciprocal,Penrose_1955} of $C_{jk}$ to reduce the noise, i.e., we carry out the eigenvalue decomposition for the covariance matrix and ignore the eigenvalues less than $10^{-3}\lambda_{max}$, where $\lambda_{max}$ is the largest eigenvalues.

As reported in \cite{HSCY3cosmos}, There are non-negligible redshift shifts in large photo-z bins. Thus we also introduce redshift shift parameter $\Delta z_{4,5}$ for the largest two redshift bins to model the uncertainty of redshift distribution:
\begin{equation}
    n_{4,5}(z)\to n_{4,5}(z-\Delta z_{4,5})
\end{equation}
Note that we use the notation that a positive $\Delta z$ means a higher true redshift. The prior ranges are both $[-0.3,0.3]$.

In total, we have two cosmological parameters, one IA parameter and two photo-z systematic parameters. All parameters are applied with flat priors. The cosmological parameters include matter density $\Omega_m$ and a combination parameter of matter density and matter fluctuation $S_8=\sigma_8\sqrt{\Omega_m/0.3}$. Other cosmological parameters are fixed with the results of Plank18 \citep{planck18}. The range of $\Omega_m,S_8$ is $[0.1,0.6],[0.5,1.0]$ separately. The IA parameter $A_{IA}$ is from NLA model. And the prior ranges are $[-6,6]$. 
We use the python package emcee \footnote{https://github.com/dfm/emcee}\citep{emcee} to perform the MCMC sampling, which implements the affine-invariant ensemble sampler for MCMC. We run 32 walkers with $10^4$ steps to sample each posterior distribution.

\begin{figure}[htbp]
    \centering
    \includegraphics[width=\textwidth]{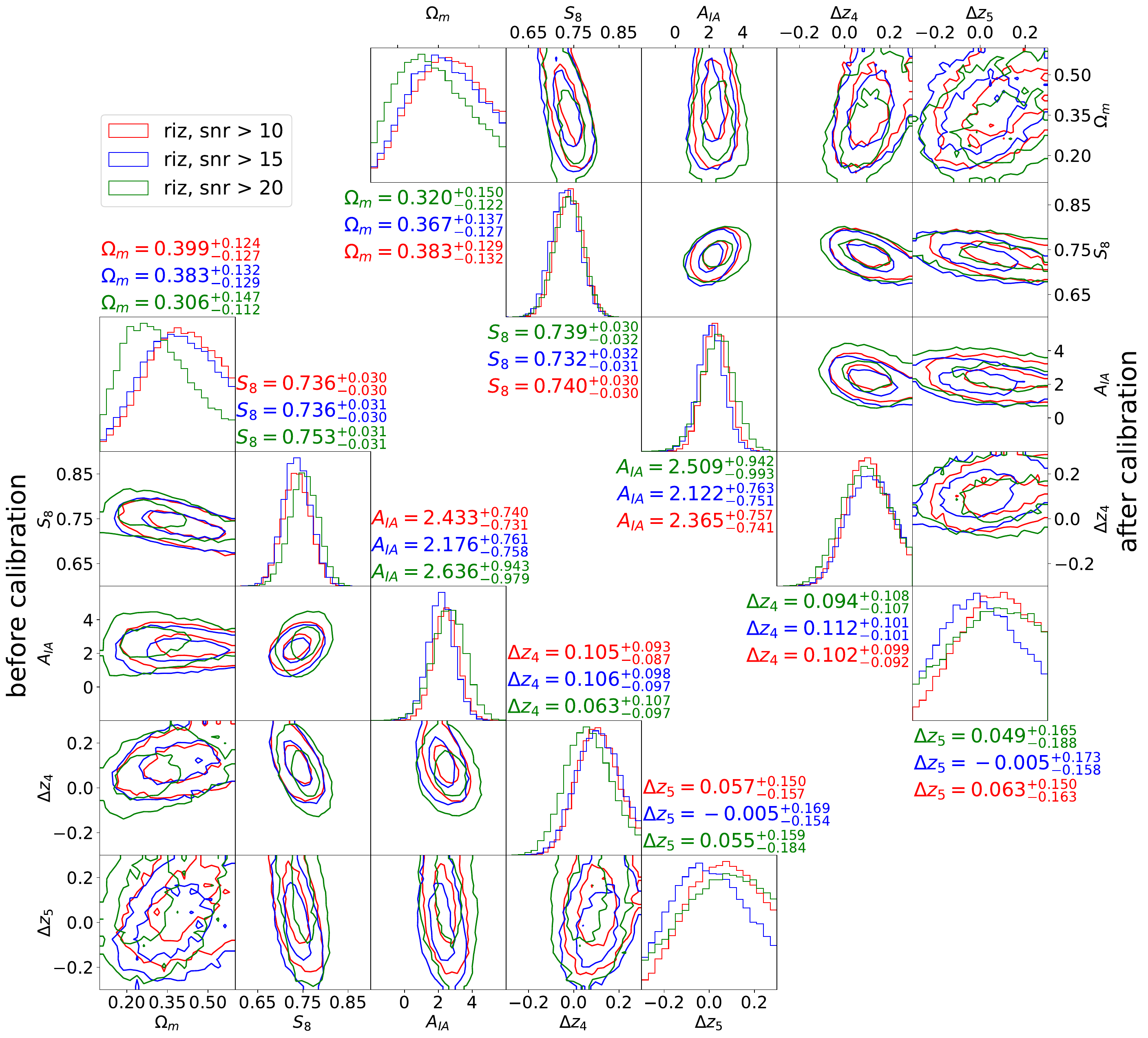}
    \caption{The marginalized posteriors distributions of all free parameters ($\Omega_m,S_8,A_{IA},\Delta z_{4,5}$). The bottom left panels represent results from the uncalibrated 2PCFs of r/i/z bands under various SNR cuts using FQ estimators. And the top right panels represent the results from the calibrated 2PCFs. Different colors in each panel represent different SNR selections. The 2D contours represent the 68\% and 95\% confidence levels. The histograms represent the marginalized 1D posterior distributions. The best fit values are marked on the top of the bottom-left histograms and bottom of the top-right histograms. }
    \label{fig:mcmc_riz}
\end{figure}

The final constraint contours of r/i/z bands under different SNR cuts using FQ estimators are shown in Fig.\ref{fig:mcmc_riz}. In this Figure, we can compare the contours before and after onsite calibration. We find that the parameters (espically $\Omega_m,S_8$) are more consistent with each other after calibration. Considering the constraint power, we choose the results from r/i/z bands shear catalog with $\mathrm{SNR}>10$ cut as our fiducial results, which gives $\Omega_m=0.383^{+0.129}_{-0.132}$, $S_8=0.740^{+0.030}_{-0.030}$. The IA amplitude is constrained to be $A_{IA}=2.365^{+0.757}_{-0.741}$. And due to the large errorbars, we do not find significant redshift shift, which are constrained to be $\Delta z_4=0.102^{+0.099}_{-0.082}$, $\Delta z_5=0.063^{+0.150}_{-0.163}$.

\begin{figure}[htbp]
    \centering
    \includegraphics[width=\textwidth]{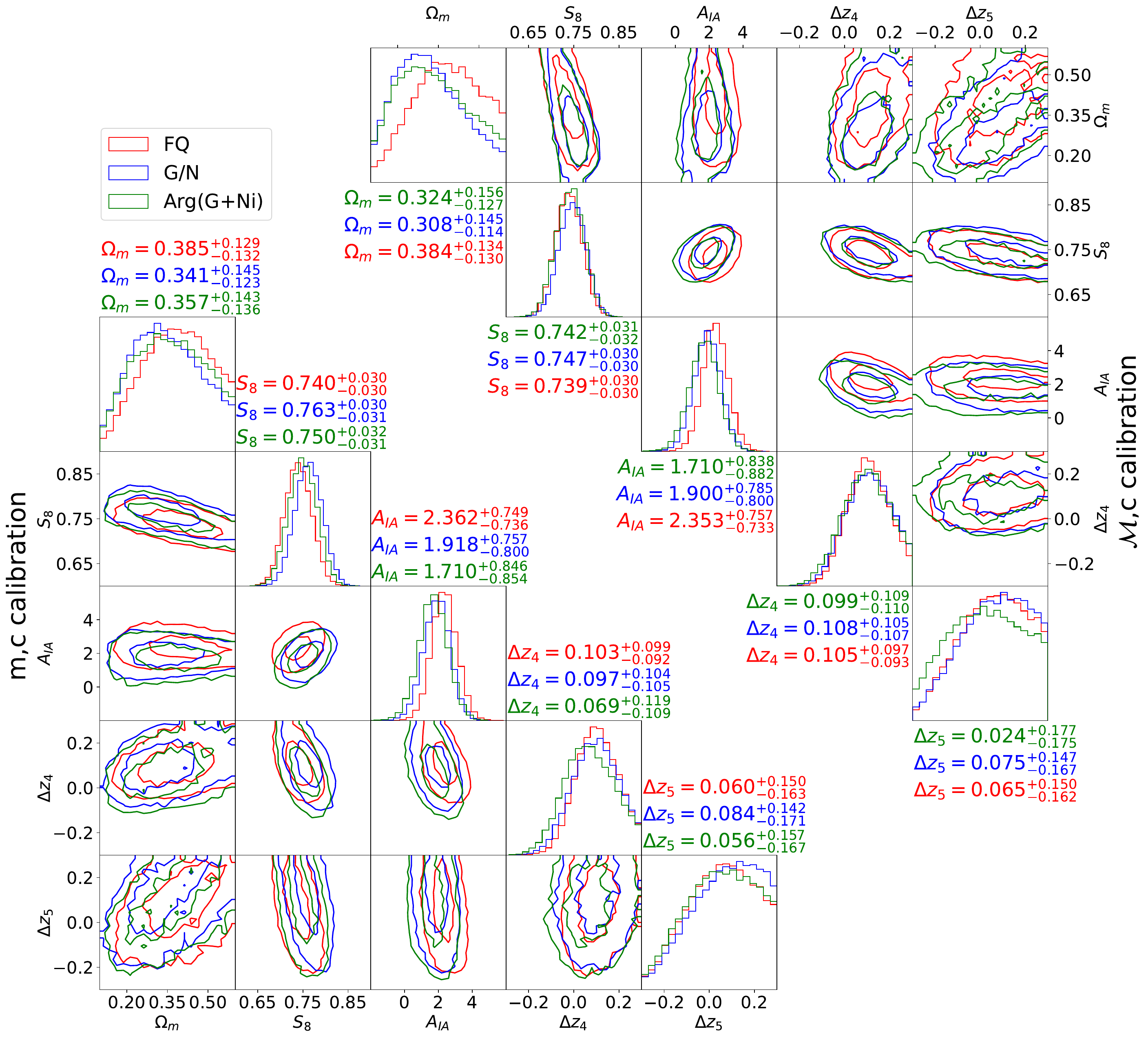}
    \caption{Similar to Fig.\ref{fig:mcmc_riz}, but for different shear estimators, all with $\mathrm{SNR}>10$ cut. In this Figure, we show the posterior distribution from the 2PCFs calibrated with m,c and $\mathcal{M},c$ in the bottom-left and the top-right panels respectively.}
    \label{fig:mcmc_gn}
\end{figure}

In the top right panels of Figure \ref{fig:mcmc_gn}, we show the constraint contours of different shear estimators with $\mathrm{SNR}>10$ cut. Note that as mentioned in section \ref{sec:shear_cat}, we decide to use $\mathcal{M},c$ to calibrate the 2PCFs for $\mathrm{G/N}$ and $\mathrm{Arg(G+Ni)}$ shear estimators. Thus we also plot the contours calibrated by m from FD test in the bottom left panels of this figure. We find that the final parameters constraint (especially $S_8$) are more consistent with each other when calibrated with $\mathcal{M},c$. These results confirm our selection of the calibration method. It also indicates that the correlation of the stochastic shear bias for certain shear estimators are not likely negligible in general.

We also measure the 2PCFs and constrain the cosmology using i band only catalog. The results show similar properties, and are also consistent with those from the r/i/z bands. The errorbars of 2PCFs and constraint parameters from i band are $\sim20\%$ larger than riz bands, indicating there are extra information from the multi-band shear measurements. This is consistent with our previous work \citep{FQ_HSC}.

\section{conclusion \& discussion }
\label{sec:conclusion}

In weak lensing shear statistics, galaxies are usually selected with certain cuts to ensure the quality of shear estimators. These selections may introduce selection biases, which should be carefully calibrated to avoid biased results. In this paper, we raise the idea of onsite calibration, which directly estimates and calibrates the biases for every specific selection on the real shear catalog. We apply this idea to the FQ shear catalog from  HSCpdr3 images. We measure the 2PCFs and constrain cosmological parameters with different selections. The results show consistency after calibration.

We use both the FD test  and the recently proposed TC to estimate the multiplicative and additive shear biases. We find that the biases vary with different SNR cuts, photo-z bins, bands, and shear estimators. This indicates that the shear biases depend on galaxy properties, and thus confirms the necessity of onsite calibration. We also find that for FQ shear estimators, the m from FD test and $\mathcal{M}$ from TC are generally consistent. But for $\mathrm{G/N}$ and $\mathrm{Arg(G+Ni)}$ shear estimators, m and $\mathcal{M}$ show deviations, which indicates that these two estimators may have different responses to spatially correlated residuals. Thus, we choose to use m,c to calibrate the FQ shear estimators, and $\mathcal{M},c$ to calibrate the $\mathrm{G/N}$ and $\mathrm{Arg(G+Ni)}$ shear estimators.

We measure the shear shear correlation functions with five tomographic redshift bins. We vary different SNR cuts, bands and shear estimators to test the consistency of 2PCFs. In Fig \ref{fig:2pcf_riz},\ref{fig:2pcf_gn} we show that 2PCFs are consistent with each other after onsite calibration. 

We further carry out cosmology constrains with the measured 2PCFs. The final constrained parameters (especially $S_8$ ) again show consistency under various selections. Considering the constraining power, we choose the FQ results from the r/i/z bands shear catalog with $\mathrm{SNR}>10$ cut as our fiducial results, which gives $\Omega_m=0.383^{+0.129}_{-0.132}$, $S_8=0.740^{+0.030}_{-0.030}$. 

\begin{figure}[htbp]
    \centering
    \includegraphics[width=0.5\textwidth]{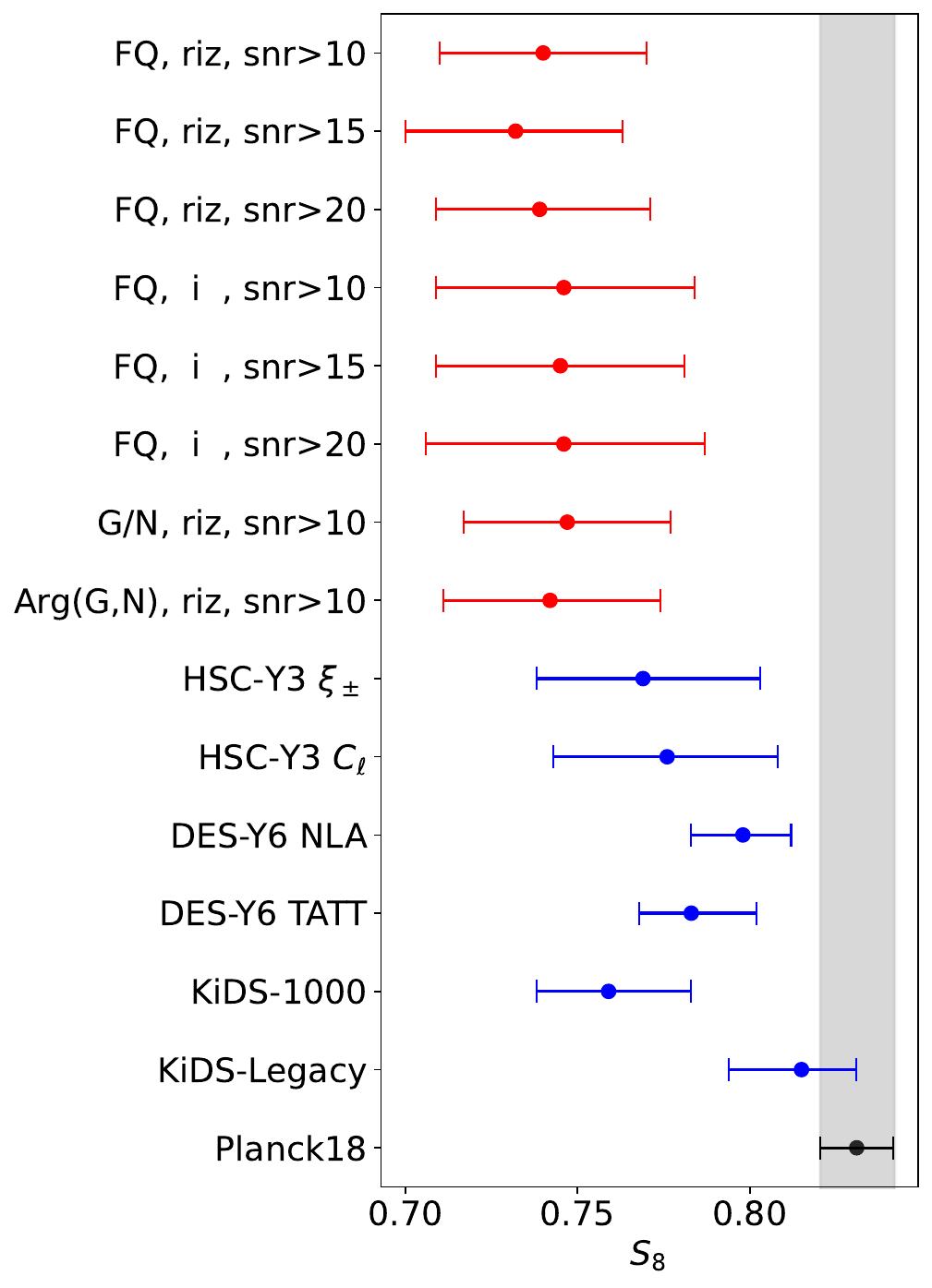}
    \caption{Summary of the $S_8$ constraints from various cuts and shear estimators, and also from other works.}
    \label{fig:s8_compare}
\end{figure}

We summarize our $S_8$ constraints in Fig.\ref{fig:s8_compare}. We also present several external results from weak lensing surveys, e.g., the fiducial 2PCF analysis from DES-Y6 \citep{DESy6cosmos}, the fiducial 2PCF and shear power spectra results from HSC-Y3 \citep{HSCY3cosmos,HSCY3cosmos2}, and the fiducial COSEBIs results from  KiDS-1000 \citep{KiDS1000cosmos} and KiDS-Legacy \citep{KiDs_Legacy}. The fiducial constraints from Planck18 \citep{planck18}  Cosmological Microwave Background (CMB) results are also included. We find that our $S_8$ constraints from various selections and shear estimators are generally consistent with each other. Compared to other weak lensing results, our results are also consistent with DES-Y6, HSC-Y3 and KiDS-1000 results, which are systematically lower than Planck18 with $\sim 2\sigma$.

However, KiDS-Legacy shows consistency with Planck18. As proposed in \cite{KiDs_Legacy}, This difference mainly comes from the careful calibration of redshift distribution. As mentioned in Section \ref{sec:2pcf}, our treatment for redshift distribution is rather simple. In this work, we mainly focus on the 2PCF measurement part, the detailed study of redshift distribution calibration is left for future works.

As shown in Figure \ref{fig:2pcf_riz},\ref{fig:2pcf_gn}, our fiducial small scale cut is $\sim$ 7 arcmin, where the baryonic effects are not significant. Thus, we do not include baryonic effects correction in our theoretical model. We also note that both the large scale cut and the number of data points are limited by jackknife method. We will explore other covariance matrix estimation methods to extend the scale ranges, and include smaller scale information using baryonic corrections.

\normalem
\begin{acknowledgements}
  This work is supported by the National Key Basic Research and Development Program of China (2023YFA1607800, 2023YFA1607802), the NSFC grants ((12573004), and the science research grants from China Manned Space Project (No. CMS-CSST-2021-A01). The computations in this paper are run on the $\pi$ 2.0 and the Siyuan-1 cluster supported by the Center of High Performance Computing at Shanghai Jiao Tong University, and the Gravity supercomputer of the Astronomy Department, Shanghai Jiao Tong University.
\end{acknowledgements}

\bibliographystyle{raa}
\bibliography{thesis}

\end{document}